\documentclass[11pt]{article}
\usepackage{amsfonts}
\usepackage{dsfont}
\usepackage{amssymb}
\usepackage{amsmath}
\usepackage{amsxtra}
\usepackage{graphicx}
\usepackage{psfrag}
\usepackage{mathrsfs}
\usepackage{multirow}
\usepackage{stmaryrd}
\usepackage{subfigure}
\usepackage{empheq}
\usepackage[normalem]{ulem}
\usepackage{pdfpages}
\usepackage{wrapfig}
\usepackage{longtable}
\usepackage{footmisc}
\usepackage{hyperref}
\usepackage{lineno}
\usepackage{outlines}
\usepackage{enumitem}
\usepackage{lipsum}
\usepackage{upgreek}
\usepackage{bm}
\usepackage{scalerel}

\usepackage{float}
\usepackage{authblk}
\usepackage{caption} 
\usepackage[margin=0.75in]{geometry}
\usepackage{cite}
\usepackage{soul}
\usepackage{dblfloatfix}
\usepackage{siunitx}
\usepackage{tikz}
\usepackage{sectsty}
\subsectionfont{\fontsize{14}{15}\selectfont}

\newcommand{\jbe}[1]{\textcolor{gray}{\textbf{(jbe)} #1}}

\newcommand{\tb}[1]{\textcolor{black}{#1}}

\newcommand{\ignore}[1]{}

\newcommand{\up}[1]{\mathbf{#1}}

\begin{document}
\def\BGamma{\mbox{\boldmath$\Gamma$}}
\def\BDelta{\mbox{\boldmath$\Delta$}}
\def\BTheta{\mbox{\boldmath$\Theta$}}
\def\Btheta{\mbox{\boldmath$\theta$}}
\def\BLambda{\mbox{\boldmath$\Lambda$}}
\def\BXi{\mbox{\boldmath$\Xi$}}
\def\BPi{\mbox{\boldmath$\Pi$}}
\def\BSigma{\mbox{\boldmath$\Sigma$}}
\def\BUpsilon{\mbox{\boldmath$\Upsilon$}}
\def\BPhi{\mbox{\boldmath$\Phi$}}
\def\BPsi{\mbox{\boldmath$\Psi$}}
\def\Btheta{\mbox{\boldmath$\Omega$}}
\def\Balpha{\mbox{\boldmath$\alpha$}}
\def\Bbeta{\mbox{\boldmath$\beta$}}
\def\Bgamma{\mbox{\boldmath$\gamma$}}
\def\Bdelta{\mbox{\boldmath$\delta$}}
\def\Bepsilon{\mbox{\boldmath$\epsilon$}}
\def\Bzeta{\mbox{\boldmath$\zeta$}}
\def\Beta{\mbox{\boldmath$\eta$}}
\def\Btheta{\mbox{\boldmath$\theta$}}
\def\Biota{\mbox{\boldmath$\iota$}}
\def\Bkappa{\mbox{\boldmath$\kappa$}}
\def\Blambda{\mbox{\boldmath$\lambda$}}
\def\Bmu{\mbox{\boldmath$\mu$}}
\def\Bnu{\mbox{\boldmath$\nu$}}
\def\Bxi{\mbox{\boldmath$\xi$}}
\def\Bpi{\mbox{\boldmath$\pi$}}
\def\Brho{\mbox{\boldmath$\rho$}}
\def\Bsigma{\mbox{\boldmath$\sigma$}}
\def\Btau{\mbox{\boldmath$\tau$}}
\def\Bupsilon{\mbox{\boldmath$\upsilon$}}
\def\Bphi{\mbox{\boldmath$\phi$}}
\def\Bchi{\mbox{\boldmath$\chi$}}
\def\Bpsi{\mbox{\boldmath$\psi$}}
\def\Bomega{\mbox{\boldmath$\omega$}}
\def\Bvarepsilon{\mbox{\boldmath$\varepsilon$}}
\def\Bvartheta{\mbox{\boldmath$\vartheta$}}
\def\Bvarpi{\mbox{\boldmath$\varpi$}}
\def\Bvarrho{\mbox{\boldmath$\varrho$}}
\def\Bvarsigma{\mbox{\boldmath$\varsigma$}}
\def\Bvarphi{\mbox{\boldmath$\varphi$}}
\def\bone{\mbox{\boldmath$1$}}
\def\bzero{\mbox{\boldmath$0$}}
\def\bnabla{\mbox{\boldmath$\nabla$}}
\def\bvarepsilon{\mbox{\boldmath$\varepsilon$}}
\def\bA{\mbox{\boldmath$ A$}}
\def\bB{\mbox{\boldmath$ B$}}
\def\bC{\mbox{\boldmath$ C$}}
\def\bD{\mbox{\boldmath$ D$}}
\def\bE{\mbox{\boldmath$ E$}}
\def\bF{\mbox{\boldmath$ F$}}
\def\bG{\mbox{\boldmath$ G$}}
\def\bH{\mbox{\boldmath$ H$}}
\def\bI{\mbox{\boldmath$ I$}}
\def\bJ{\mbox{\boldmath$ J$}}
\def\bK{\mbox{\boldmath$ K$}}
\def\bL{\mbox{\boldmath$ L$}}
\def\bM{\mbox{\boldmath$ M$}}
\def\bN{\mbox{\boldmath$ N$}}
\def\bO{\mbox{\boldmath$ O$}}
\def\bP{\mbox{\boldmath$ P$}}
\def\bQ{\mbox{\boldmath$ Q$}}
\def\bR{\mbox{\boldmath$ R$}}
\def\bS{\mbox{\boldmath$ S$}}
\def\bT{\mbox{\boldmath$ T$}}
\def\bU{\mbox{\boldmath$ U$}}
\def\bV{\mbox{\boldmath$ V$}}
\def\bW{\mbox{\boldmath$ W$}}
\def\bX{\mbox{\boldmath$ X$}}
\def\bY{\mbox{\boldmath$ Y$}}
\def\bZ{\mbox{\boldmath$ Z$}}
\def\ba{\mbox{\boldmath$ a$}}
\def\bb{\mbox{\boldmath$ b$}}
\def\bc{\mbox{\boldmath$ c$}}
\def\bd{\mbox{\boldmath$ d$}}
\def\be{\mbox{\boldmath$ e$}}
\def\bff{\mbox{\boldmath$ f$}}
\def\bg{\mbox{\boldmath$ g$}}
\def\bh{\mbox{\boldmath$ h$}}
\def\bi{\mbox{\boldmath$ i$}}
\def\bj{\mbox{\boldmath$ j$}}
\def\bk{\mbox{\boldmath$ k$}}
\def\bl{\mbox{\boldmath$ l$}}
\def\bmm{\mbox{\boldmath$ m$}}
\def\bn{\mbox{\boldmath$ n$}}
\def\bo{\mbox{\boldmath$ o$}}
\def\bp{\mbox{\boldmath$ p$}}
\def\bq{\mbox{\boldmath$ q$}}
\def\br{\mbox{\boldmath$ r$}}
\def\bs{\mbox{\boldmath$ s$}}
\def\bt{\mbox{\boldmath$ t$}}
\def\bu{\mbox{\boldmath$ u$}}
\def\bv{\mbox{\boldmath$ v$}}
\def\bw{\mbox{\boldmath$ w$}}
\def\bx{\mbox{\boldmath$ x$}}
\def\by{\mbox{\boldmath$ y$}}
\def\bz{\mbox{\boldmath$ z$}}
\def\shrug{\texttt{\raisebox{0.75em}{\char`\_}\char`\\\char`\_\kern-0.5ex(\kern-0.25ex\raisebox{0.25ex}{\rotatebox{45}{\raisebox{-.75ex}"\kern-1.5ex\rotatebox{-90})}}\kern-0.5ex)\kern-0.5ex\char`\_/\raisebox{0.75em}{\char`\_}}}
\numberwithin{equation}{section}
\def\checkmark{\tikz\fill[scale=0.4](0,.35) -- (.25,0) -- (1,.7) -- (.25,.15) -- cycle;} 


\title{Ogden Material Calibration via Magnetic Resonance Cartography, Parameter Sensitivity, and Variational System Identification}

\author[a]{Denislav~P.~Nikolov}
\author[a]{Siddhartha~Srivastava}
\author[a]{Bachir~A.~Abeid}
\author[a,b]{Ulrich~M.~Scheven}
\author[a,b,c]{Ellen~M.~Arruda}
\author[a,d,e]{Krishna~Garikipati}
\author[a,*]{Jonathan~B.~Estrada}

\affil[a]{Department of Mechanical Engineering, University of Michigan}
\affil[b]{Department of Biomedical Engineering, University of Michigan}
\affil[c]{Program in Macromolecular Science and Engineering, University of Michigan}
\affil[d]{Department of Mathematics, University of Michigan}
\affil[e]{Michigan Institute for Computational Discovery and Eng., University of Michigan}
\affil[*]{To whom correspondence should be addressed. E-mail: jbestrad@umich.edu}

\maketitle

\subsection*{Abstract}
Contemporary material characterisation techniques that leverage deformation fields and the weak form of the equilibrium equations face challenges in the numerical solution procedure of the inverse characterisation problem. 
As material models and descriptions differ, so too must the approaches for identifying parameters and their corresponding mechanisms.
The widely-used Ogden material model can be comprised of a chosen number of terms of the same mathematical form, which presents challenges of parsimonious representation, interpretability, and stability. 
Robust techniques for system identification of any material model are important to assess and improve experimental design, in addition to their centrality to forward computations.
Using fully 3D displacement fields acquired in silicone elastomers with our recently-developed magnetic resonance cartography (MR-\textit{\textbf{u}}) technique on the order of $>20,000$ points per sample, we leverage PDE-constrained optimisation as the basis of variational system identification of our material parameters. 
We incorporate the statistical F-test to maintain parsimony of representation. 
Using a new, local deformation decomposition locally into mixtures of biaxial and uniaxial tensile states, we evaluate experiments based on an analytical sensitivity metric, and discuss the implications for experimental design. 
\subsection*{Keywords}
continuum mechanics $\mid$ magnetic resonance $\mid$ sensitivity $\mid$ full-field deformations $\mid$ physics inference

\section{Introduction}

Mechanical characterisation is a common area of interest for all classes of materials. 
In the most fundamental sense, there are three essential components to common experimental characterisation: (1) kinematic information, usually as grip-to-grip distances or surface data using 2D/3D digital image correlation, (2) force/nominal stress information, usually measured by load cell(s), and (3) a constitutive relation linking kinematics and stress for the material at hand. 
Frequently, the experimental procedure requires assumptions to be made about the through-thickness motion, which makes this general pipeline rather tenuous in its application to more complex sample geometries such as in biological tissues. 
These restrictive assumptions also contribute to a need for multiple testing geometries, where, e.g. uniaxial versus biaxial behaviour can be distinguished. 
However, choosing a constitutive model for the behaviour of biological tissues is non-trivial, due to the tremendous variety of different models which may or may not capture the relevant physics of the material. 
While some models are based on \tb{tensor invariants of functions of the deformation gradient tensor}, others, such as Prof. RW Ogden's eponymous model~\cite{Ogden1972, Ogden1972b} and variants thereof~\cite{Holzapfel2000}, are excellent at fitting real experimental data but are \tb{instead formulated in terms of principal stretches}. 

The implications of the mathematical model forms on our characterisation of materials are notable. \ignore{R2C6}
Techniques that leverage the weak form of mechanical equilibrium---namely, those that employ measurement of full-field deformations, such as the virtual fields method~\cite{VFMBook2012}, finite element model updating~\cite{Kavanagh1971}, or variational system identification as in our own work~\cite{Wang2021} \tb{can usefully leverage richness in this information}. 
\tb{Rather than averaging, for example, a heterogeneous strain field at many load steps to generate an average stress-strain curve, kinematic variations instead provide an avenue for one-load characterisation---i.e., using one entire boundary value problem for material calibration.} 
\tb{However, to distinguish the experimental usefulness of one kinematic field from another, we must have some sense of how the material behaves in response to deformation and a way to assess this \textit{sensitivity}.} 
\tb{Unfortunately, the parameterisation of many constitutive models in terms of the tensor invariants of the left (or right) Cauchy-Green tensor $I_i$ complicates this effort, as these} invariants are neither decoupled from one another, nor of obvious use in describing a test of a general material model not formulated in terms of these invariants. 
Further, the inverse problem of material characterisation in general may not be unique, particularly if model terms contain equivalent mathematical forms. 
\tb{Thus, we seek a framework that allows us to not only assess a material by leveraging full-field information, but \textit{also inform how good of a job we are doing} in our assessment.} 

\tb{The ability to probe full deformation fields has recently instigated several computational attempts to use the aforementioned kinematic richness for inference of constitutive and structural properties of materials.
Most notably, the development of model-free data-driven approaches}~\cite{Kirchdoerfer2016, Ibanez2017, Nguyen2018, Conti2018} \tb{surrogate constitutive models}~\cite{Ghaboussi1991, Sussman2009, Crespo2017, Zhang2020} \tb{have encountered success and have yielded extrapolative models for simulating material response.
However, the lack of interpretable model parameters in terms of the functional form of constitutive laws restricts the utility of these approaches for system inference of known physical contributions.
Another family of approaches aims to minimise the error between simulation and data using instantiations of governing equations for specified physical mechanisms (i.e., constitutive laws). 
Due to the usage of the weak form of partial differential equations (PDEs) in these approaches, we refer to these approaches as variational system identification (VSI).
A subclass of these approaches, as presented in our prior work}~\cite{Wang2021, Wang2019, Wang2020} \tb{and Flaschel et al.}~\cite{Flaschel2021}\tb{, identify the constitutive parameters that minimise the error in the  residual of the weak form of the PDE evaluated on the data fields.
This VSI framework, for the special case in which the model can be expanded in terms of a basis with parameters only appearing as coefficients of these basis terms, reduces to a quadratic unconstrained minimisation problem that is inexpensive to solve in comparison to the general non-linear optimisation problem.
This feature makes this framework especially attractive for identifying relevant terms from a large pool of admissible mechanisms.
However, this favorable structure of quadratic optimisation again becomes a non-linear optimisation problem when these parameters no longer appear linearly in the description of the model.
We note that these approaches do not rely on the forward solve of the PDE and only use the data field to minimise the residual.  
Therefore a good performance of the trained model is usually not guaranteed in the forward solution.}

\tb{
Meanwhile, PDE-constrained minimisation using the weak form, another VSI approach, utilises the forward solution of the PDE to train the model, thereby guaranteeing the quality of inferred parameters.
In this framework, model parameters are iteratively chosen using an optimisation algorithm that minimises the difference between data and simulation.
Typically these optimisation schemes (e.g., finite element model updating; FEMU}~\cite{Clough1971}\tb{), rely on gradient-based techniques such as the Gauss-Newton approach, wherein the gradient is estimated using methods like forward differences and input perturbations.
These iterative optimisation techniques usually scale linearly with the number of parameters. 
An alternative approach is an adjoint-based method that can be used to estimate parameter updates with a single forward solve}~\cite{Farrell2013, dolfin-adjoint}.
\tb{The adjoint-based method has been successfully applied to problems in optimal control, PDE-constrained optimisation, topology optimisation and sensitivity analysis.
PDE-constrained optimisation is especially useful for inferring PDEs with fewer parameters, which is beneficial when the model is phenomenological.}
\begin{figure}[b!]
    \centering
    \includegraphics[width=\textwidth]{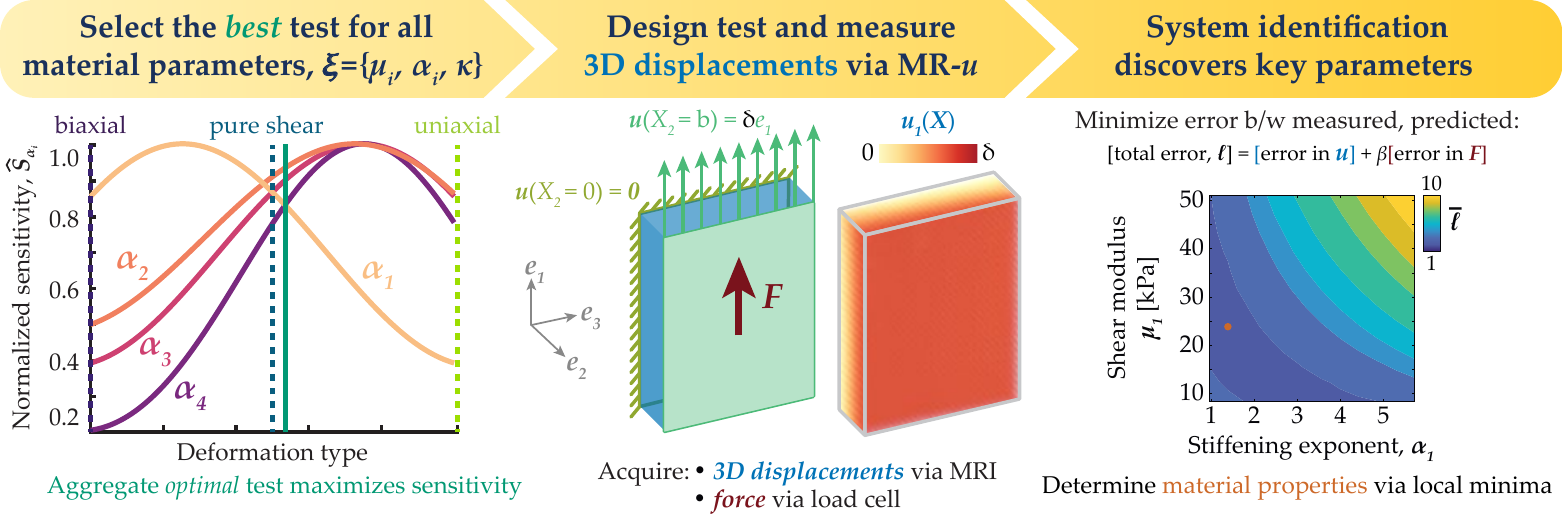}
     \caption{
    \tb{\textbf{Flowchart for test design and material calibration.} After choosing a material model, an experiment is evaluated on how well it will predict, or how sensitive it is to, a set of material parameters based on sensitivity metrics constructed from the kinematics and constitutive behaviour.
    We design a test to maximise the aggregated experimental sensitivity, and then minimise a kinematic+stress-based error metric, $\ell$ to determine the best-fit material properties.}}
    \label{fig:abstract}
\end{figure}

In this work, \tb{we present a methodology for choosing an experiment for hyperelastic characterisation, combining it with PDE-constrained variational system identification, and refining it using an experimental sensitivity metric (illustrated in flowchart form in figure}~\ref{fig:abstract}\tb{).}  
\tb{We leverage our recent experimental development of alternating pulsed field gradient stimulated echo imaging (APGSTEi)}~\cite{Scheven2020}\tb{, which provides the full 3D deformation field between a reference and an actuated deformed, or mapped, state; we describe this mapping technique as magnetic resonance cartography (MR-\textit{\textbf{u}})}~\cite{Estrada2020}.
Notably, MR-\textit{\textbf{u}} acquires fully three-dimensional displacement fields without addition of fiducial speckle patterns as in digital image correlation~\cite{Chu1985} or internal contrast~\cite{Bay1999}/fluorescent particles as in digital volume correlation~\cite{Franck2007, Stout2016}. 
We distinguish two styles of double-lap shear experiments for silicone rubbers at finite strains, with differing degrees of heterogeneity by inclusion of cylindrical holes. 
Using a linear motor actuator with a displacement encoder in series with a load cell, we measure loads concurrently with the acquired deformation fields. 
\tb{We couple our full-field methods with a deformation orthogonalisation (or mode-mixture) approach, parameterised by a stretch magnitude and a uniaxial-vs-biaxial parameter in similar spirit to the approach of Criscione}~\cite{Criscione2000, Criscione2001}. 
We then analytically link the type and amplitude of the local deformation field to a local sensitivity of our constitutive function to deformation, and in turn, an overall experimental goodness metric. 
We conclude with a discussion on the evaluation of experiments using these metrics, with a long-term goal of procedurally optimising experimental geometries for material calibration using a given model.

\section{Theory and Methods}\label{sec:theory}

We start by considering an arbitrary solid body $\mathcal{B}$ with volume $\Omega_0$ in its reference configuration, denoted by internal coordinates $\bX$. 
Upon application of prescribed displacements $\bu^*(\bX \in \partial \Omega_0)$ and tractions $\bp^*(\bX \in \partial \Omega_0)$ on the surface $\partial \Omega_0$, the body $\mathcal{B}$ is assumed to undergo a finite deformation into a new configuration given by a volume $\Omega_i$ and new mapped coordinates $\bx$. 
The non-translational portion of the mapping is described by the deformation gradient tensor  $\up{F}(\bX) = \bnabla_{\bX} \bx(\bX) = \bnabla_{\bX} \bu(\bX) + \up{I}$, where $\bu(\bX)=\bx(\bX)-\bX$ is the vector displacement field between the reference and deformed states and $\up{I}$ is the second-order identity tensor. 
The experiments in this paper use our previously described pulse sequence and apparatus with a different sample geometry~\cite{Scheven2020, Estrada2020} to extract $2\pi$-wrapped displacement fields using nuclear magnetic resonance (NMR). 
We define other kinematic quantities of interest using the deformation gradient tensor, such as the left and right Cauchy-Green tensors $\up{B}=\up{F}\up{F}^\intercal$ and $\up{C}=\up{F}^\intercal \up{F}$, respectively, and the right stretch tensor $\up{U}=\sqrt{\up{C}}$. 
We write the right stretch tensor in spectral form as $\up{U}= \sum_{j=1}^3 \lambda_j \bn^{(j)} \otimes \bn^{(j)}$, from the right polar decomposition of the deformation gradient tensor, $\up{F}=\up{R}\up{U}$, where $\lambda_j$ represent the three principal stretches of the deformation \tb{in their respective orthogonal eigenframe directions, $\bn^{(j)}$}. 
While the object is acted upon in general by tractions $\bp^*$, practical experimental limitations (i.e., measurement with load cells) define the measured quantity as the axial component $F_{\textrm{data}}$ of the total load $\bP$ acting on a subsurface $\Gamma_{p^*}$ of the object surface $\partial \Omega_0$, where
\begin{equation}
    \bP = \int_{\Gamma_{p^*}} \bp^*(\bX) dS.
\end{equation}

\subsection{Material Model and Sensitivity Metrics}
Our prior investigation of platinum-cure silicone rubber focused on our fidelity in the extraction of material parameters either (1) assuming specific hyperelastic functions~\cite{Estrada2020} while employing the virtual fields method \tb{(VFM)}~\cite{VFMBook2012, Promma2009} of Pierron, Grediac, et al. or (2) remaining agnostic of the exact form of constitutive properties~\cite{Wang2021} while employing \tb{VSI}~\cite{Wang2019}. 
To date, our efforts have focused on \tb{tensor} invariant-based constitutive laws. 
It is thus only natural---and in this special issue, a joy!---to extend this approach to include RW Ogden's eschewal of \tb{the tensor invariants of the left or right Cauchy-Green tensors} for his eponymous material model~\cite{Ogden1972}, given by the isochoric free energy potential
\begin{equation}
\Psi_{\textrm{Ogden}} = \displaystyle \sum\limits_{i=1}^{N} \frac{\mu_i}{\alpha_i} \left(\lambda_1^{\alpha_i}+\lambda_2^{\alpha_i}+\lambda_3^{\alpha_i} - 3 \right).     
\end{equation}
The material parameters $\mu_i$ and $\alpha_i$ represent corresponding shear moduli and stiffening exponent coefficients for the separate terms of $\Psi_{\textrm{Ogden}}$, and must agree with the shear modulus at small strain $\mu$, by the relation $2\mu = \sum_i^N \mu_i\alpha_i$. 
The choices of $\mu_i$ and $\alpha_i$ are not entirely independent; to satisfy Hill's stability criterion, the products $\mu_i\alpha_i>0$ is necessary and sufficient for the strain energy to be positive-definite in the case of two or fewer terms included in the model~\cite{Ogden1972}. 
For three or more terms, these conditions are sufficient, but not all of these conditions are necessary; stability can be checked using e.g. the approach of Johnson~\cite{Johnson1994}. 

\begin{figure}[b!]
    \includegraphics[width=\textwidth]{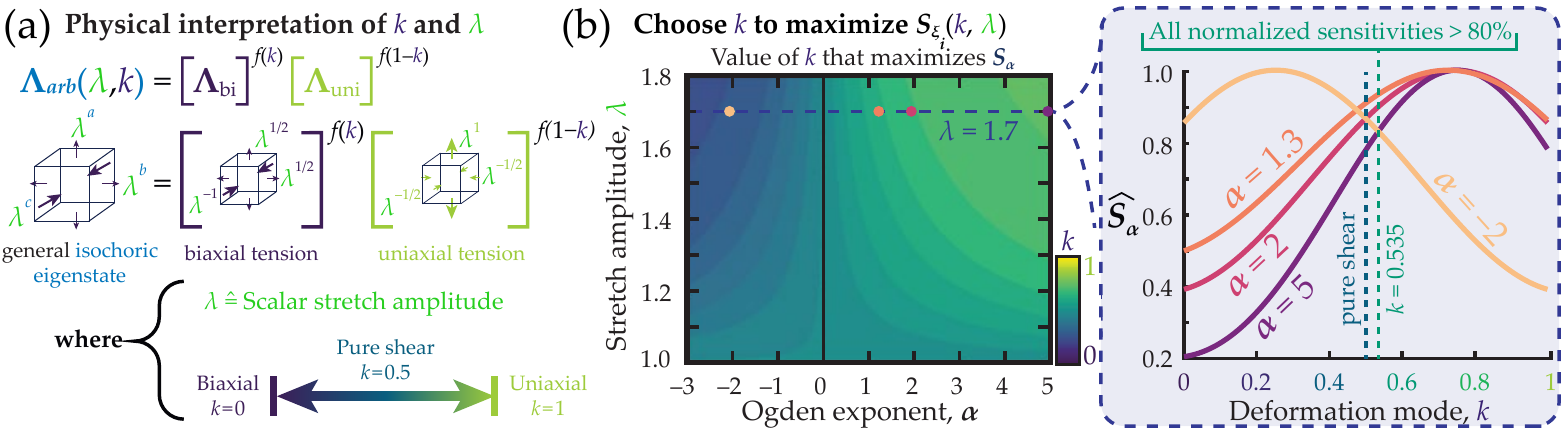}
     \caption{
     \textbf{Selection of double-lap shear test is guided by sensitivity.} 
     (\textit{a}) By decomposing the isochoric part of an arbitrary deformation into a mixture of biaxial and uniaxial tensile modes, we can 
     (\textit{b}) optimise the choice of experiment by the expected sensitivity for estimating different parameters. 
     At a stretch of 1.7 \tb{(blue dashed line, left and plot, right)}, the value of $k=0.535$ maximises the \tb{normalised} sensitivity \tb{for the alpha parameter, $\hat{S}_\alpha$,} for the domain of $-2<\alpha<5$, suggesting shear-like tests to be optimal. 
     }
    \label{fig:sensitivty}
\end{figure}

Before proceeding to parameter fits, we first present a metric for quantifying the expected sensitivity of an experiment to the perceived correct values of $\mu_i$ and $\alpha_i$. 
We use a modified analytical version of a discrete \tb{local sensitivity metric} presented by Marek et al.~\cite{Marek2017,Marek2019} \tb{in the context of the VFM}.
\tb{This metric was used previously for identifying material properties governing plastic deformation behaviour,} which required a time-differential form.
\tb{In our} hyperelastic context \tb{we use the metric analytically to} guide the kind of experiment we should run to maximise our signal-to-noise ratio on our parameter estimates. 
\tb{We define this goodness metric as our sensitivity, and define it as a function of two parameters: the type of deformation denoted by a parameter $k$ on $[0,1]$, where 0 corresponds to biaxial tension, and 1 corresponds to uniaxial tension), and a scalar amplitude of the stretch state we denote as $\lambda$.}
\tb{The decomposition of deformation is similar in spirit to the physically interpretable invariants presented by Criscione}~\cite{Criscione2000, Criscione2001}\tb{, with an important distinction that we employ a formulation that permits use of work functions of the deformation gradient tensor or associated principal stretches, as opposed to functions only of logarithmic strain. }
To define our sensitivity to a change in a material parameter, which we intend to be the amount of energy change with respect to both a change in applied stretch and a desired parameter, we first (in a nod to RW Ogden) move away from \tb{tensor} invariant-based deformation formulations by ``orthogonalising'' the constitutive deformations themselves. 
We start by treating the isochoric and volumetric portions of the deformation independently, in accordance with separate terms in the free energy potential. 
We decompose an arbitrary deformation $\up{F}_{\textrm{arb}}$\tb{$=J^{1/3}\bar{\up{F}}_{\textrm{arb}}$} into a volumetric part $J^{1/3} \up{I}$ and isochoric part $\bar{\up{F}}_{\textrm{arb}}$, where $J= \det(\up{F}_{\textrm{arb}})$.
Then, we rotate the corresponding right stretch tensor to $\bar{\up{F}}_{\textrm{arb}}T$, $\bar{\up{U}}_{\textrm{arb}}$, into its eigenframe via $\bm{\Uplambda}_{\textrm{arb}} = \up{Q} \bar{\up{U}}_{\textrm{arb}} \up{Q}^\intercal$, where $\up{Q}$ is the change-of-basis tensor from the eigenframe. 
We may then describe the three eigenvalues of $\bm{\Uplambda}_{\textrm{arb}}$ using a multiplicative composition of isochoric biaxial and uniaxial tension,
\begin{equation}
    \bm{\Uplambda}_{\textrm{arb}} = \bm{\Uplambda}_{\textrm{uni}}^{f(k)} \bm{\Uplambda}_{\textrm{bi}}^{f(1-k)} = \begin{bmatrix}
\lambda  & 0 & 0\\
0 & \lambda^{-1/2} & 0 \\
0 & 0 & \lambda^{-1/2}
\end{bmatrix}^{f(k)}
\begin{bmatrix}
\lambda^{1/2}  & 0 & 0\\
0 & \lambda^{1/2} & 0 \\
0 & 0 & 1/\lambda
\end{bmatrix}^{f(1-k)}, 
\end{equation}
where $f(k)$ is a mixture function of these two constitutive deformation states (figure~\ref{fig:sensitivty}\textit{a}). 
Adding compressibility, or volumetric expansion/contraction, to $\bm{\Uplambda}_{\textrm{arb}}$ would just require multiplication by $J^{1/3}$, which could be similarly parameterised by a separate coefficient $m$, i.e. having $J^{1/3} = \lambda^m$. 
To ensure that the case of pure shear yields eigenvalues of $\{\lambda, 1, \lambda^{-1}\}$ at $k=0.5$, \tb{and uniaxial extension holds for $k=1$ and equibiaxial extension holds for $k=0$}, a simple satisfactory polynomial function is $f(k) = (-2k^2+5k)/3$, thus parameterising the three eigenvalues of the stretch in terms of $\lambda$ and $k$ as
\begin{equation}
    \lambda_1 = \lambda^{-k^2+\frac{3}{2}k+\frac{1}{2}}, ~~\lambda_2 = \lambda^{-k+\frac{1}{2}}, ~~\lambda_3 = \lambda^{k^2-\frac{1}{2}k-1 }.
\end{equation}
For brevity, we define three values $\bg(k)=\{g_1=-k^2+\frac{3}{2}k+\frac{1}{2}, g_2=-k+\frac{1}{2}, g_3=k^2-\frac{1}{2}k-1\}$. 
Note that, as expected, the relation $\lambda_1\lambda_2\lambda_3=1$ is \tb{automatically} satisfied in accordance with isochoric deformation. 
From here we cast the isochoric part of the Ogden free energy potential itself as a function of $(\lambda,k)$, 
\begin{equation}
    \Psi_\textrm{Ogden}(\lambda,k) = \displaystyle \sum\limits_{i=1}^{N} \frac{\mu_i}{\alpha_i} \left(\lambda^{\alpha_i g_1}+\lambda^{\alpha_i g_2}+\lambda^{\alpha_i g_3} - 3 \right).
\end{equation}
If we elect to include limited compressibility, the strain-energy function of the Ogden model can be defined with a bulk term as
\begin{equation}\label{eq:comp_Ogden}
    \Psi = \Psi_{\textrm{Ogden}}\left(\overline{\lambda}_1, \overline{\lambda}_2, \overline{\lambda}_3  \right) + \frac{1}{2} \kappa \left(\ln{J} \right)^2, 
\end{equation}
where $\overline{\lambda}_i= J^{-1/3}\lambda_i$ represent the isochoric versions of the principal stretches. 
The first Piola-Kirchhoff stress $\bm{\Uppi} = d\Psi(\up{F})/d\up{F}$ is then 
\begin{align}
    \bm{\Uppi} &= \left( \sum_{j=1}^3 \tau_j \bn^j \otimes \bn^j  \right) \up{F}^{-\intercal}, \\
    & \text{where } \tau_j = \displaystyle \sum\limits_{i=1}^{N} \mu_i J^{-\alpha_i/3}\left(\lambda_j^{\alpha_i} - \frac{1}{3} \left( \lambda_1^{\alpha_i} + \lambda_2^{\alpha_i} + \lambda_3^{\alpha_i} \right) \right) + \kappa \ln J.
\end{align}

We are now interested in the shape of the isochoric free energy function landscape with respect to the stretch ``magnitude'' $\lambda$ and the deformation mixture parameter $k$, where $k=0$ represents isochoric biaxial tension, $k=0.5$ represents a state of pure shear, and $k=1$ represents isochoric uniaxial tension, respectively (as visualised in figure~\ref{fig:sensitivty}\textit{a}). 
Using an analytical scalar form of the sensitivity metric inspired by that presented in Marek et al.~\cite{Marek2017,Marek2019} and another used in our own prior work~\cite{Luetkemeyer2021}, we define the sensitivity of $\Psi$ to a material parameter $\xi_i$---i.e., how much the slope of the work-stretch response $\partial \Psi/\partial \lambda$ changes based on perturbations in our material parameter,
\begin{equation}
    S_{\xi_i} = \frac{\partial^2 \Psi_{\textrm{Ogden}}}{\partial \xi_i \partial \lambda} = \frac{\partial \bm{\Uppi}}{\partial \xi_i} \bm{:} \frac{\partial \up{F}}{\partial \lambda},
\end{equation}
which can alternately be written in terms of the first Piola-Kirchhoff stress.
\tb{The sensitivity metrics $S_{\xi_i}$ describe the change of this stress function with small changes in a given material parameter $\xi_i$ while under an experimental deformation state $\up{F}$. 
For an isochoric deformation state, we would expect the bulk term sensitivity function $S_\kappa$ to be zero, for example.} 
We may now cast \ignore{our sensitivity metrics} $S_{\xi_i}(\lambda,k,\bm{\xi})$ \tb{analytically}, where $\bm{\xi}$ represents our vector of material parameters. 
As each term in the Ogden potential is assumed to be linearly independent of the others, partial derivatives with respect to parameters $\xi_i$ also eliminate the summation, leaving the sensitivity functions $S_{\mu_i}$ and $S_{\alpha_i}$ 
\begin{equation}
    S_{\mu_i}(\lambda, k, \bm{\xi}) = g_1 \lambda^{\alpha_i g_1-1}+g_2 \lambda^{\alpha_i g_2-1}+g_3 \lambda^{\alpha_i g_3-1}
\end{equation}
\begin{equation}
    S_{\alpha_i}(\lambda, k, \bm{\xi}) = \mu_i \ln\lambda\left[g_1^2 \lambda^{\alpha_i g_1-1}+g_2^2 \lambda^{\alpha_i g_2-1}+g_3^2 \lambda^{\alpha_i g_3-1}\right].
\end{equation}

Figure~\ref{fig:sensitivty}\textit{b} shows how maximising $S_{\alpha_i}$ with respect to $k$ will reveal what states of stretch are most suitable to particular $\alpha_i$ terms, and how for a chosen subset of terms, a single simple shear test is determined to be the best candidate experiment for the $\alpha_i$ terms traditionally associated with rubber-like material~\cite{Treloar1943}.


\subsection{Magnetic Resonance-Displacement Field Acquisition} \label{mracqoutline}

\begin{figure}[t!]
    \includegraphics[width=\textwidth]{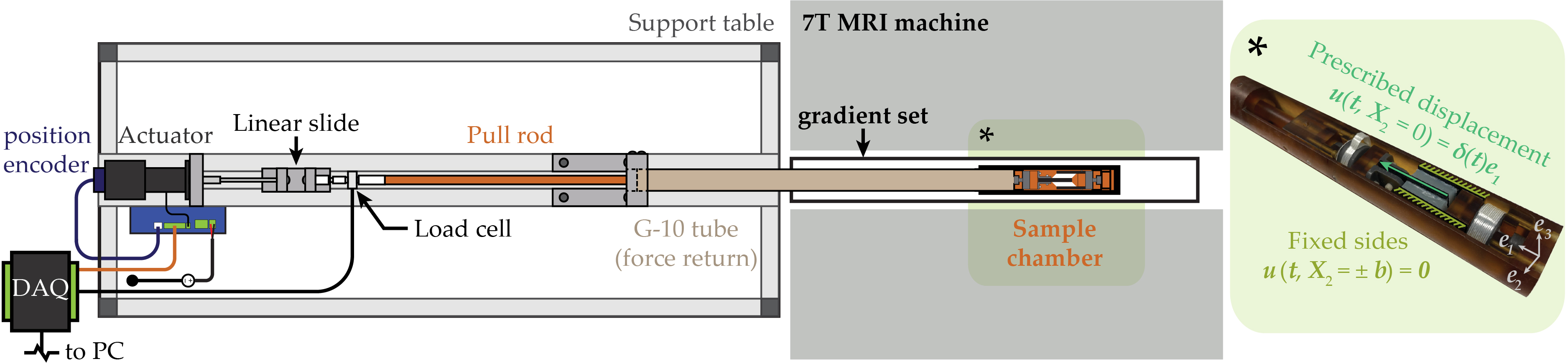}
     \caption{
    \textbf{Schematic of the magnetic resonance cartography (MR-\textbf{\textit{u}}) setup.} A confined linear actuator in series with a load cell supplies a nominal displacement to a lap shear sample via a pull-rod. The sample is placed at the center of a \SI{7}{\tesla} 
    magnetic resonance imaging machine, and is cyclically loaded in a custom chamber (star, on right) between its reference and deformed states by a prescribed global displacement.
     }
    \label{fig:MRISchematic}
\end{figure}

The procedure for acquiring full, 3D displacement fields in samples undergoing large deformations has been presented in our prior work, with details on the pulse sequence~\cite{Scheven2020} and image processing/coil correction~\cite{Estrada2020} procedures, as well as a brief overview in~\cite{Wang2021}. 
For the ease of the reader, we briefly describe here the core methods of, as well as relevant modifications to, the technique. 
As shown in figure~\ref{fig:MRISchematic}, a polyetherimide chamber for material testing in global uniaxial deformation is situated in the center of a \SI{7}{\tesla} 
small-animal MRI system. 
The chamber is mounted to a glass-reinforced composite tube and connected with a concentric polyetherimide pull-rod controlled by a distant captive linear actuator (L5918S2008-T10X2-A50, Nanotec Electronic GmbH and Co. KG, Germany). 
The actuator supplies a known global displacement profile, and is placed far enough away from the MRI machine that it does not appreciably interact with the strong magnetic field. 
Informed by the sensitivity metric definition, we designed and fabricated custom grips to create an approximate double-lap shear experiment of two cast rectangular silicone samples (Ecoflex OO-20 formulation; Dragon Skin, Smooth-On Inc., Macungie, PA) glued using cyanoacrylate to comparatively rigid 1/4" polyacrylic plates. 
Two types of sample were tested: solid rectangular, and rectangular with two cylindrical holes per sample. 
For samples containing inclusions, the cylindrical holes were produced by fixing cylinders onto the base of the mold and cutting off the residual material with a circular punch of the same corresponding size.

We induced approximate simple shear in the silicone samples and imaged them simultaneously and synchronously with the MRI. 
A load cell (LCM300, Futek Advanced Sensor Technology Inc., Irvine, CA) in series with the pull-rod and samples measured the total axial force applied by the actuator; load accuracy was additionally verified using an external uniaxial testing device. 
\tb{We obtained images of the displacements and strains using data acquired with the APGSTEi MRI pulse sequence described in}~\cite{Scheven2020}. 
\tb{The sequence belongs to a larger class of displacement encoding sequences conceptually related to standard diffusion-weighted imaging. 
It employs spin-echo encoding and decoding blocks to eliminate field distortion artefacts that might otherwise arise in MRI at \SI{7}{\tesla}. 
Spatial encoding and decoding segments of the sequence are typically separated by an interval $\Delta$ = \SI{400}{\milli\second} 
during which time the sample is taken from the strained state to the reference state. 
After the decoding block has run, a standard imaging readout provides the data for image reconstruction. 
The images are complex-valued, and the phase of each pixel is proportional to the displacement - along the chosen displacement encoding direction - which the protons in that pixel have undergone as the sample was taken from the stretched state to the undeformed reference state.
Data are acquired for all three orthogonal displacement encoding directions in order to obtain the full displacement and strain fields.
These phase maps can then be unwrapped to yield displacement maps by a user-preferred method}~\cite{Jenkinson2002, AbdulRahman2007}, \tb{or numerically differentiated to construct the deformation gradient tensor $\up{F}$ by direct division of adjacent pixels, or by incorporating image processing kernel differentiation (or \textit{divolution}) as in our prior work}~\cite{Estrada2020}.
\tb{In this work, we chose to use direct numerical differentiation as in}~\cite{Scheven2020} \tb{to suppress the introduction of unwrapping artefacts that may arise from larger divolution filters. 
While we expect our future efforts in experimental optimisation to investigate this in detail, our method proposed herein used the undifferentiated displacement field in all but calculating experimental sensitivity.}    

\subsection{System Identification}\label{subsec:VSI}

In this section, we present a systematic approach to construct a minimal Ogden model that describes full-field displacement acquired from MR-\textit{\textbf{u}} experiments.
We use PDE-constrained optimisation techniques that have been widely applied for parameter estimation for PDEs~\cite{Wang2021,Wang2019,Wang2020,dolfin-adjoint,Elouneg2021}. 
The variational formulation used in the PDE-constrained optimisation is presented first with a subsequent discussion on the inference methodology.



\subsubsection{Variational Formulation and PDE-constrained Optimisation}

\tb{Variational system identification is a class of techniques that falls under the broader umbrella of model inference.
It is related to strong form-based approaches~\cite{KutzPNAS2015,KutzSCIADV2017,KutzSIAM2019}, but enjoys advantages of smoothness conferred by the weak form, which is important for noisy data.}
The weak form of the stress equilibrium equation is written as
\begin{equation}\label{eq:weakform}
\int_\Omega \frac{\partial \boldsymbol{w}}{\partial\boldsymbol{X}}\colon \bm{\Uppi}~\mathrm{d}V - \int_{\Gamma_{p^*}} \boldsymbol{w}\cdot\boldsymbol{p^*}~\mathrm{d}S = 0,
\end{equation}
where $\boldsymbol{w}$ represents the weighting function (an arbitrary variation on the displacement field). 
Following the standard procedure in finite element analysis (FEA) computations of discretising the domain and accounting for the arbitrariness of the weighting function's degrees of freedom, equation (\ref{eq:weakform}) leads to a set of nonlinear algebraic equations written as the residual vector, $\bR = \bzero$. 
A more rigorous discussion on the theory and implementation of the non-linear finite element formulation of such problems are presented in our earlier work~\cite{Wang2021}. 

In a typical formulation for PDE-constrained optimisation for parameter estimation of models, the error between the deformation field estimated using the forward solution and experimental displacement data is minimised. 
However, this optimisation is ill-posed for a displacement controlled experiment; i.e., for a displacement-driven boundary value problem: The absence of the boundary traction in (\ref{eq:weakform}) leaves the linear coefficients undetermined up to a common multiplicative constant. Therefore, since the displacement-controlled experiments performed in this work also provide the net force applied on the surface, the \tb{error} is augmented to include the difference in predicted $\int_{\Gamma_{p^*}} \bp^* \cdot\bn~ \text{d}S$ and observed $F_{\textrm{data}}$ values of total force on the boundaries: 
\begin{align}
    \lbrace\kappa, \alpha_1, \mu_1 \cdots, \alpha_N, \mu_N   \rbrace= 
    \underset{\lbrace\widetilde{\kappa} , \widetilde{\alpha_1}, \widetilde{\mu_1} \cdots, \widetilde{\alpha_N}, \widetilde{\mu_N}  \rbrace}{\text{arg }\min} \text{ } \left\{ \ell_u + \beta \ell_F \right\}
\end{align}
\begin{align*}    
    \text{where } \ell_u = \frac{1}{\Omega} \int\limits_{\Omega} \frac{\vert\bu^\text{FE}-\bu \vert^2}{|\bu|_{max}^2}\text{d}V \qquad \ell_F = \left(1 - \frac{1}{F_{\textrm{data}}}\int\limits_{\Gamma_{p^*}} \bp^* \cdot\bn ~\text{d}S\right)^2\nonumber\\
    \text{subject to: } \boldsymbol{R}=\bzero,
\end{align*}
where $\bu^\text{FE}$ is the displacement field from the forward FEA solution obtained with the current values of coefficients  $\lbrace\kappa, \alpha_1, \mu_1 \cdots, \alpha_N, \mu_N   \rbrace$ and $\bu$ is the corresponding experimental full-field displacement data \tb{with maximum absolute value denoted as $|\bu|_{max}$. The term $\ell_u $ represents the error in the displacement data, while $\ell_F$ represents the error in the measured traction data on the loading boundary. The total error is defined as a weighted mean of these two error components parameterised with a user-defined scale factor $\beta$.}
We adopt a standard implementation of the adjoint method to evaluate the gradient, which requires a single solve of the (linear) adjoint equation of the original PDE constraint and a nonlinear solution for the parameter set. 
In this work we use the Sequential Least Squares Programming \tb{(SLSQP)} optimisation algorithm from the \texttt{SciPy} package~\cite{2020SciPy} and the \texttt{dolfin-adjoint} software library~\cite{dolfin-adjoint} to solve the PDE-constrained minimisation problem.

\subsubsection{Parsimony in Operator Selection}
\begin{figure}[t!]
    \includegraphics[width=\textwidth]{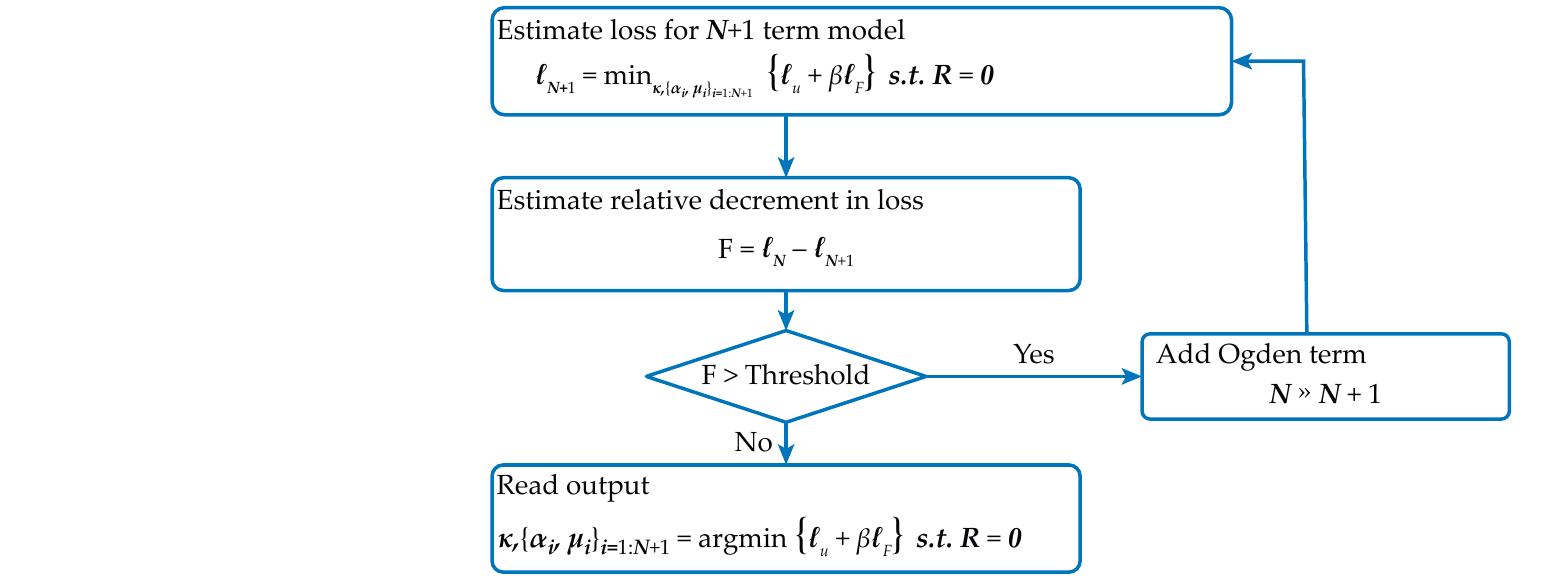}
     \caption{\textbf{Procedural flowchart for constructing a parsimonious Ogden model.}
     }
    \label{fig:algo_ftest}
\end{figure}

As is often the case with regression techniques, having a large number of terms may lead to  over-fitting. 
Therefore, it is desirable to train the Ogden model with a minimal number of terms. 
To achieve this, we will start with a one-term model and successively introduce more terms; keeping additional terms only if the resulting decrements in \tb{error} are above a certain threshold value. 
This statistical criterion is often referred to as the $F$-test. 
We have also used it in our previous work~\cite{Wang2019,Wang2020} for basis reduction in system identification problems. 
The significance of the change in the model by addition of a term in an $N$-term model is evaluated as:
\begin{align}
\text{F-value}= \ell_{N} - \ell_{N+1}
\end{align}
where $\ell_N$ is the \tb{error} of the best-fit $N$-term Ogden model. If the F-value falls below the threshold, there is not a significant improvement in augmenting the $N$-term model with an additional term. However, as long as the F-value remains above the threshold, the addition of terms improves the model.
A procedure outlining the algorithm for choosing a parsimonious model is presented in figure~\ref{fig:algo_ftest}.

\subsubsection{Numerical implementation and stability constraints}

The phenomenology of the Ogden model makes it challenging to train an interpretative model, for instance, constructing distinct terms that allow for different mechanisms of deformation, such as strain-stiffening. 
Moreover, the Ogden model is not stable for all choices of the parameter set, and therefore some stability conditions are required on the parameter space for training.
Typically, a 2-3 term model is sufficient for fitting materials~\cite{yeoh1989phenomenological}.
However, in principle there is no restriction on the number of terms. 
It is therefore crucial to use statistics to determine if the proposed model is over-fitting the data. 

Hill's stability criteria \tb{is} required as a necessary condition and can be readily imposed as a quadratic inequality constrained in terms of training parameters that most optimisation packages are well-poised to solve, for instance~\cite{2020SciPy} and~\cite{diamond2016cvxpy}. 
\tb{We chose to use \texttt{SciPy}}~\cite{2020SciPy} \tb{in this work.} 
Problematically, equivalence in the form of each term of the Ogden model introduces degeneracy in the \tb{error metric}, for instance, (1) the different Ogden terms can be permuted without changing the model, and (2) the model response is insensitive to $\mu_i$ in the regime $\alpha_i \rightarrow 0$, and similarly to $\alpha_i$ when $\mu_i \rightarrow 0$. 
The first degeneracy \tb{is} eliminated by considering an ordered sequence of $\alpha_1 \geq \cdots \geq \alpha_i \geq \cdots \geq \alpha_{N}$. 
Noting that it is desirable to have some separation in the $\alpha$ values, we consider the following inequality constraint, with $\eta = 1$ in our simulations:
\begin{equation}\label{eq:min_gapa_const}
    \alpha_{i} > \alpha_{i+1} + \eta.
\end{equation}
From a numerical implementation standpoint, it \tb{is} more efficient to formulate the problem in terms of the parameters $(\kappa, \alpha_1, \mu_1, \cdots, \Delta \alpha_i = \alpha_i - \alpha_{i-1}, \mu_i, \cdots  )$. 
Here we replaced the parameter $\alpha_i$ with $\Delta \alpha_i$ which allow\tb{ed} us to implement Eq~\eqref{eq:min_gapa_const} as bounds, in contrast to equality constraints, that are often more efficiently handled by the solver due to their specialised form. 
All the simulations for PDE-constrained optimisation presented here \tb{were} carried out with this new parameter set; however, the results are presented in the traditional format. 
We lastly impose\tb{d} a condition on the shear moduli of $ \sum_{i=1}^N \mu_i > 0$ to ensure Hill's stability of the material. 

\section{Results}\label{sec:results}

\subsection{Full-field displacement data}
\begin{figure}[t!]
    \includegraphics[width=\textwidth]{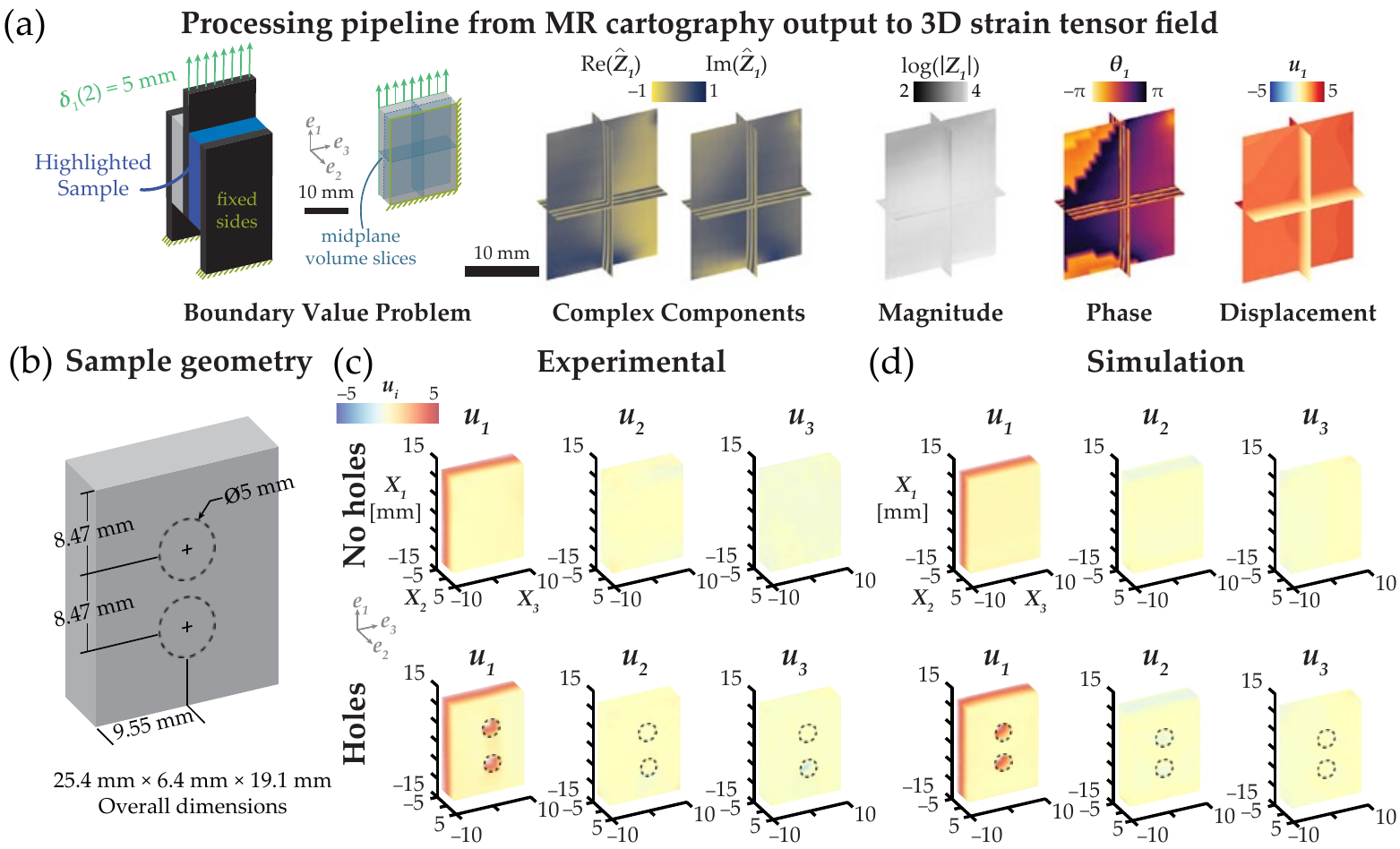}
     \caption{
    \textbf{Schematic of processing pipeline and 3D plots of displacement fields.}
    (\textit{a}) The complex, fully 3D raw data extracted from the double lap shear test is converted to magnitudes and phases, the latter of which define displacement fields. 
    (\textit{b}) \tb{Upon fabricating the appropriate sample containing no holes and containing two holes (dotted black line),
    the displacement fields} are determined by numerical differentiation, (\textit{c}) which compare favorably with simulations.
     }
    \label{fig:processing}
\end{figure}

Samples were loaded in a double-lap shear configuration, shown schematically in figure~\ref{fig:processing}\textit{a}, and stretched to three prescribed global displacement levels ([2.5, 5, 7] \SI{}{\milli\meter} over the thickness of \SI{6.4}{\milli\meter}).
Two types of samples were tested;
the outer dimensions of all silicone samples were 25.4$\times$6.4$\times$19.1 \SI{}{\cubic\milli\meter}.
\tb{The second type of sample contains two \SI{5}{\milli\meter} diameter holes spaced out evenly in the longitudinal direction and positioned symmetrically on the center-line of the sample, with detailed dimensions shown in figure}~\ref{fig:processing}\tb{\textit{b}.}
Displacement fields were acquired with the procedure \tb{summarised in section}~\ref{mracqoutline} \tb{and illustrated in figure}~\ref{fig:processing}\tb{\textit{a}}.
\tb{Figure}~\ref{fig:processing}\tb{\textit{a} shows an unwrapped phase map proportional to the local displacement along the 1-direction.
Such displacement maps were acquired for each of the three displacement directions, and yield displacement fields} via numerical differentiation; figure~\ref{fig:processing}\textit{\tb{c}} shows the \tb{displacement field} components $u_{i}(\bX)$ with an applied $2 \times 2 \times 2$ Hamming smoothing kernel~\cite{Scheven2020, Estrada2020}.
At the chosen voxel resolution of $\chi_{res} = [1,\,0.1875,\,0.75]$\SI{}{\milli\meter}, each sample has approximately 22,000 volumetric data points per prescribed global displacement, where each voxel contains the full displacement field.
Notably, the experimental displacement fields of the rectangular samples (figure~\ref{fig:processing}\textit{\tb{c}}) compare well with Abaqus FEA (Providence, RI) simulations (figure~\ref{fig:processing}\textit{\tb{d}}) of the same prescribed global displacement conditions for an approximate material.
Additionally, two samples \tb{were} tested simultaneously, resulting in both experimental symmetry and two full-field data sets. 

\begin{figure}[t!]
    \includegraphics[width=\textwidth]{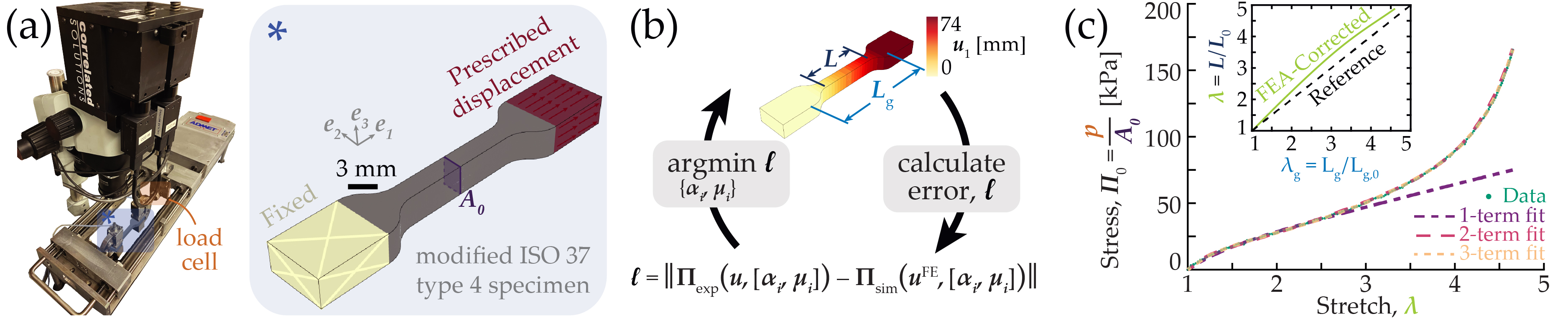}
     \caption{
    \textbf{Uniaxial testing validation procedure.} (\textit{a}) A uniaxial testing frame (ADMET, Norwood, MA) extends an ISO 37 Type 4 specimen (star) via a prescribed, quasistatic grip-to-grip ramp profile. (\textit{b}) The sample was rendered in Abaqus FEA and extended using prescribed rigid displacements on the dog-bone flanges, given an initial guess of material properties, and evaluated/updated in terms of \tb{an error metric} to best match experimental data. (\textit{c}) Fitting results for 1- (purple, dash-dotted), 2- (pink, dashed), and 3-term (peach, dotted) Ogden models to experimental data (blue). FEA-corrected  grip-to-grip to gauge stretch function, inset. 
     }
    \label{fig:uni_validation}
\end{figure}
\subsection{Uniaxial test validation}\label{subsec:tensile_train}

\tb{To verify our system identification method, we performed a comparative} baseline low-rate uniaxial test, as shown in figure~\ref{fig:uni_validation}\textit{a}. 
The uniaxial test defined a load and grip-to-grip displacement response, which was subsequently converted to gauge-section stretch and stress using Abaqus. 
Since the experiments were displacement controlled, the deformation field, and hence, the relation between grip-to-grip and gauge section stretches \tb{were} expected to be nearly independent of the material model; this was later verified numerically. 
We built an FEA model with an identical geometry to the experiments and a dummy viscoelastic material model (figure~\ref{fig:uni_validation}\textit{b}). 
The FEA results were then used to extract a conversion polynomial-fit relation between grip-to-grip and gauge section stretches, as illustrated in figure~\ref{fig:uni_validation}\textit{c}.

Table~\ref{table:table_trained_w_hole_old} shows the fitted Ogden material model parameters for $N$ as 1, 2, or 3 terms; the fits are shown in figure~\ref{fig:uni_validation}\textit{c}. 
\tb{We performed optimisation} in three different manners: (1) constrained and unconstrained fitting to relatively low-stretch portions of the experimental stress-stretch data ($1 \leq \lambda \leq 2$) to determine the material initial shear modulus ($\mu = \frac{1}{2}\sum_{i=1}^N \alpha_i \mu_i$) using both a Neo-Hookean model (equivalently, a single term Ogden model with a fixed $\alpha_{1} = 2$) and an unconstrained single term Ogden model,  
(2) constrained fitting to the full dataset ($1 \leq \lambda \leq 4.5$) with fixed $\alpha_{1} = 2$ and bounded initial shear for two- and three-term Ogden models, and (3) unconstrained fitting to the full dataset. 

\begin{table}[b!]
\centering
\caption{\textbf{Uniaxial load-stretch experiments of an ISO 37 type 4 sample of Ecoflex OO-20 were inversely fit to finite element simulations.}
For the first two cases (Neo-Hookean and Ogden 1-term), fits were performed only up to gauge stretch values of $\lambda=2$, or before significant strain-stiffening. 
Small-strain moduli $\mu$ and specific terms $\mu_i$ and $\alpha_i$ are shown for each case. 
All moduli are in \SI{}{\kilo\pascal}. 
}
\label{table:table_trained_w_hole_old}
\begin{tabular}{ c c c cccccc c} \hline
\ignore{\multirow{2}{*}{\jbe{Case}}} & \ignore{\multirow{2}{*}{$N$}} & \ignore{\multirow{2}{*}{$\mu$}} & \multicolumn{6}{c}{\textbf{Material Parameters}} & \ignore{\multirow{2}{*}{\jbe{LSQe}}} \\
\textbf{Case} & \textbf{Terms} & $\mu$ &  $[\mu_1$ & $\alpha_1]$ & $[\mu_2$ & $\alpha_2]$ & $[\mu_3$ & $\alpha_3]$ & \textbf{LSQe} \\ \hline \hline
$1<\lambda<2.0$ & 1 & 16.3 & 16.3 & 2 (\textit{fixed}) & - & - & - & - & 0.08  \\
 & 1 & 16.1 & 16.6 & 1.94  & - & - & - & - & 0.09 \\ \hline
 
$1<\lambda<4.5$ & 2 & 16.6 & 16.6 \ignore{24.7} & 2 (\textit{fixed}) & --3.4e--7 \ignore{6e-06} &  --10 & - & - & 1.8e--4 \\
 & 3 &  15.8 & 15.8 \ignore{5.0} & 2 (\textit{fixed}) & --5.4e--10 \ignore{1.2e-7} & --34.5 & 1.3e--2 \ignore{15} & 6.4 & 6.3e--5 \\ \hline
 
$1<\lambda<4.5$ & 1 & 14 & --4.3 \ignore{14} & --6.5 & - & - & - & - & 6.3e--3  \\
 & 2 & 24.7 & --10.0 \ignore{24.7} & --4.94 & 9.45e--7 \ignore{6e-06} & 12.7 & - & - & 1.3e--4 \\
 & 3 & 20.0 & 3.03 \ignore{5.0} & 3.3 & --7.62e--9 \ignore{1.2e-7} & --31.5 & --22.6 \ignore{15} & --1.33 & 4.3e--5 \\ \hline
 \end{tabular}
\end{table}


\subsection{Inferred parsimonious Ogden model}

\begin{table}[b!]
\centering
\caption{\textbf{Inferred Ogden models using PDE-constrained optimisation with double-lap shear experiments data \tb{for samples with holes and with $\beta = 0.0004$}.} The values denoted as $0.00$ are truncated at two decimal places but represent non-zero values. \tb{All moduli are in \SI{}{\kilo\pascal}.}} 
\label{table_vsi_param}
\begin{tabular}{c c c c cc cc cc c c}
\hline
$\delta_1$ & \textbf{Filter} & $\kappa$ & $\mu$ & $[\mu_1$ & $\alpha_1]$ & $[\mu_2$ & $\alpha_2]$ & $[\mu_3$ & $\alpha_3]$ & \tb{$\ell_N$}($\times 0.001$) & \textbf{F-values} \\ \hline \hline
$2.50$ & \textit{N} & 3.988e3 & 16.86 & 33.71 & 1.00 & - & - & - & - & 13.202 &  \\
$2.50$ & \textit{N} & 8.216e3 & 15.52 & 8.02 & 0.00 & --8.02 & --3.87 & - & - & 13.160 & 0.042
 \\
$2.50$ & \textit{N} & 8.216e3 & 15.52 & 0.00 & 1.50 & 8.08 & --0.00 & --8.02 & --3.87 & 13.160 & 0.000 \\ \hline
$5.00$ & \textit{N} & 5.894e3 & 22.14 & 44.28 & 1.00 & - & - & - & - & 35.937 &  \\
$5.00$ & \textit{N} & 6.721e3 & 16.32 & 37.21 & --0.00 & --17.74 & --1.84 & - & - & 35.847 & 0.090 \\
$5.00$ & \textit{N} & 6.863e3 & 16.06 & 0.00 & 1.00 & 20.02 & 0.00 & --16.06 & --2.00 & 35.847 & 0.000 \\ \hline
$2.50$ & \textit{Y} & 2.577e3 & 16.73 & 23.08 & 1.45 & - & - & - & - & 2.875 &  \\
$2.50$ & \textit{Y} & 2.577e3 & 16.73 & 23.08 & 1.45 & --0.92 & 0.00 & - & - & 2.875 & 0.000 \\
$2.50$ & \textit{Y} & 2.577e3 & 16.73 & 0.00 & 3.03 & 23.08 & 1.45 & --0.00 & --4.11 & 2.875 & 0.000 \\ \hline
$5.00$ & \textit{Y} & 2.173e3 & 16.24 & 17.00 & 1.91 &  - & - & - & - & 5.724 &  \\
$5.00$ & \textit{Y} & 2.190e3 & 16.18 & 17.12 & 1.89 & --4.37 &  0.00 & - & - & 5.719 & 0.005 \\
$5.00$ & \textit{Y} & 2.386e3 & 16.61 & 1.23 & 2.86 & 17.72 & 1.65 & --0.11 & --4.22 & 5.719 & 0.000 \\ \hline
\end{tabular}
\end{table}

Full-field data from individual double-lap shear experiments \tb{were} used \tb{to infer} the Ogden model with a PDE-constrained optimisation approach. 
Training \tb{was} conducted on the silicone specimens with cylindrical holes for two different prescribed boundary displacements,  $\delta = 2.5,$ \SI{5}{\milli\meter}. 
The heterogeneity in the stress and displacement fields introduced due to the holes \tb{was} necessary to generate an informative data set for inference. 
This \tb{was} in contrast to the traditional approaches, like our uniaxial testing validation approach, in which a uniform displacement field in the specimen \tb{was} desired and the dataset \tb{was} populated by taking data at increasing load steps. 
However, in this case, the data from only one single load step \tb{was} used to characterise the material. 
The training \tb{was} carried out on both raw and filtered data for 1-, 2- and 3-term Ogden models; the results are presented in table~\ref{table_vsi_param}. 
We observe\tb{d} a reduction in the \tb{error} with an increasing number of Ogden terms, an expected outcome considering that the  $N$-term model can be written as a special case of an $N+1$-term model with $\alpha_{N+1} = 0$. 
The \tb{error} in the case of unfiltered data is higher than the filtered case owing to random noise in the data. 
We note that in certain cases, $\alpha$ and $\mu$ are approximately zero. 
These terms are energetically insignificant; this is an artefact of the degeneracies that exist in the Ogden model. 

\section{Discussion}\label{sec:discussion}

\begin{figure}[b!]
    \includegraphics[width=\textwidth]{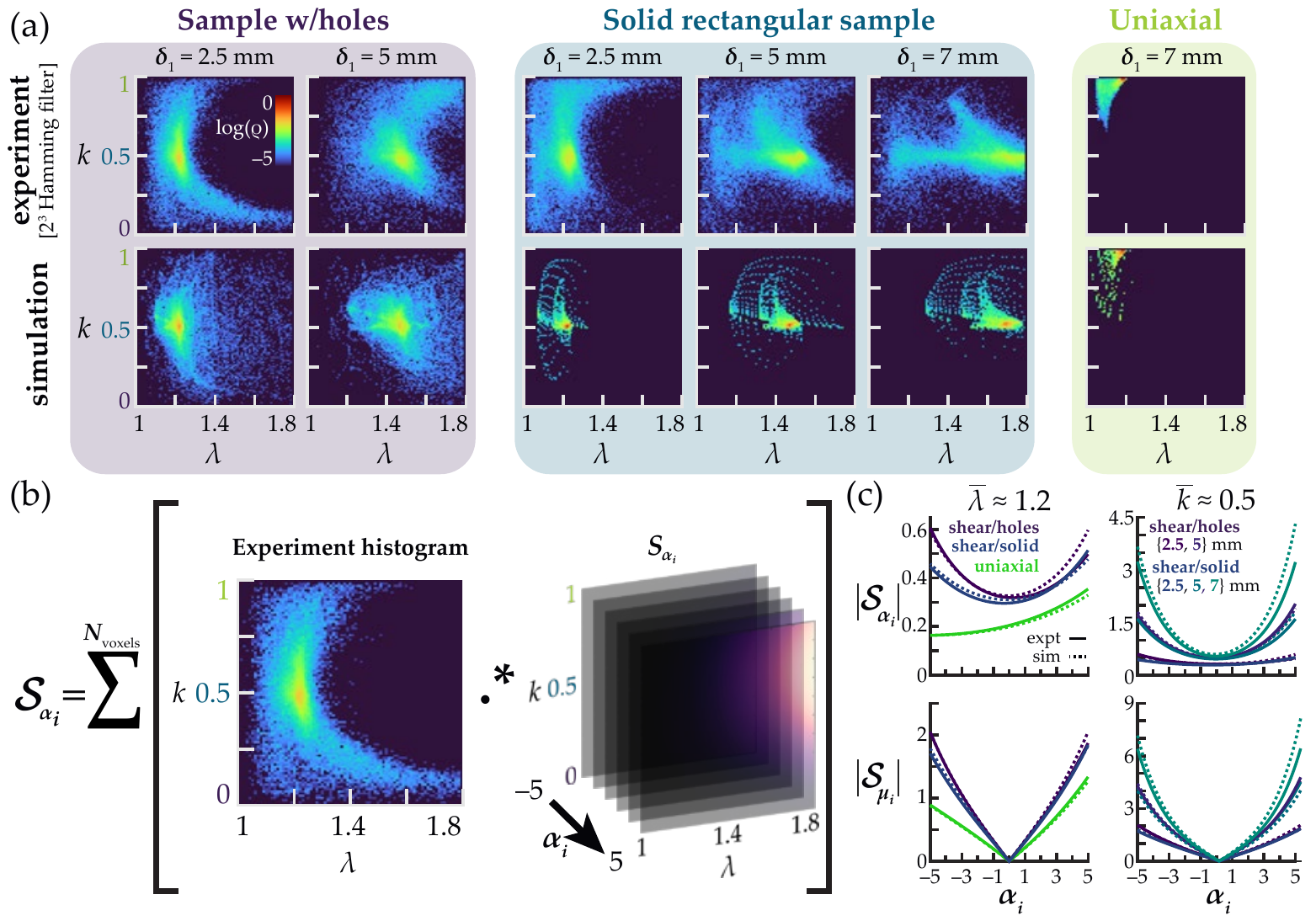}
     \caption{
    \textbf{Deformation distribution plots and implications on experimental goodness.}
    (\textit{a}) Log-density histograms of the deformation by voxel for the shear sample with holes (purple), rectangular shear sample (teal), and a comparative uniaxial set from our prior work~\cite{Estrada2020} (green) show good agreement between experimental (top row) and finite element simulation (bottom row) features. 
    (\textit{b}) An experimental goodness metric per material parameter, $\mathcal{S}_{\xi_i}$, is constructed as the discrete sum of the \tb{pointwise} product of each experimental deformation distribution with the corresponding parameter sensitivity. 
    (\textit{c}) Integrated sensitivities for $\alpha_i$ and $\mu_i$ are shown for the three geometries for cases of approximately constant average stretch (left column, \tb{$\bar{\lambda}\approx 1.2$}) and average deformation mode (right column, \tb{$\bar{k}\approx 0.5$}).
     }
    \label{fig:2D_hist_sensitivity}
\end{figure}

\subsection{Parameter sensitivity analysis}
We begin with two-dimensional log-density histograms of the decomposed deformations $\up{F}(\bX_i)$ in each of our samples at different prescribed displacement levels in figure~\ref{fig:2D_hist_sensitivity}\textit{a}. 
The vertical axis $k$ of each histogram represents the kind of isochoric eigen-deformation state experienced by a rectangular voxel in the material, where equibiaxial tension ($k=0$) and uniaxial tension ($k=1$) represent extremes, and pure shear $k=0.5$ is the midline. 
Each horizontal axis represents the scalar amplitude corresponding to this stretch state. 
The tests are grouped by style, with shear testing of EcoFlex samples containing holes (purple), shear testing of solid rectangular samples (teal), and a comparison uniaxial test \tb{(green)} of a different platinum-cure silicone (DragonSkin, SmoothOn Inc., Macungie, PA) from our prior work~\cite{Wang2021, Estrada2020}. 
The two rows denote 2D histograms---i.e., the number of voxels that have a certain leading stretch $\lambda$ and type of deformation $k$---of (i) experimental data subject to a $2\times2\times2$ Hamming filter, and (ii) finite element simulation data corresponding to each experiment, using the same approximate element count and idealised displacement conditions. 
We note several interesting characteristics of these plots. 
First, the overall voxel-deformation distribution for each experiment and its corresponding simulation look qualitatively very similar, with the locations of maxima (with values of $k\approx 0.5$ in shear and $k\approx 1$ for the uniaxial test) in very good agreement. 
Furthermore, as these data were acquired as displacement fields, numerically differentiated along each direction, and decomposed, there was significant opportunity for error due to discretisation.
This is noticeable in the density spread of the solid rectangular sample experiments. 
Even so, the principal characteristics remain the same for each plot pair. 
Inherent spread in the deformation is not unique to noise, however; the shear samples containing holes have significant point-to-point variation as confirmed by simulations; this richness is of critical importance in our identification of parameters to follow.

From the log-density histograms, we can define a new, integrated sensitivity metric $\mathcal{S}_{\xi_i}$ by taking the pointwise product of each experimental deformation density with the sensitivity to the respective material parameter $S_{\xi_i}$ and summing the result over all voxels in the sample (figure~\ref{fig:2D_hist_sensitivity}\textit{b}). 
The integrated sensitivity per parameter, which we interpret as an experimental goodness metric, illustrate improvements in our assessment of parameters by modifying the test geometry. 
In figure~\ref{fig:2D_hist_sensitivity}\textit{c}, the green curves represent the uniaxial test geometry of our prior work~\cite{Estrada2020}. 
By instead using the described simple shear geometry and controlling for median stretch (left column), $\mathcal{S}_{\xi_i}$ in all cases improves over the uniaxial test ($\sim30-200\%$) and is generally, though not universally, moderately improved by adding in heterogeneity via holes. 
Increasing the median strain but controlling for \tb{the type of} deformation (right column) improves $\mathcal{S}_{\xi_i}$ in all cases.

\subsection{PDE-constrained optimisation}

The approach of PDE-constrained optimisation offers a robust way of estimating parameters from full displacement fields \tb{in a sample with} a single prescribed \tb{edge} displacement. 
The small displacement shear modulus estimates (presented in table~\ref{table_small_strain}) match well with those of the traditional optimisation technique presented in table~\ref{table:table_trained_w_hole_old}, particularly with the cases of a fixed Neo-Hookean term. 
While the compressible variation of the Ogden model (\ref{eq:comp_Ogden}) is used in our optimisation scheme which allows for the estimation of the bulk modulus term $\kappa$,
the estimated Poisson ratio is approximately $0.5$ for all cases confirming the expected incompressible behaviour of the material. 
However, the experimental conditions must be chosen carefully to accurately capture the non-linearity in the constitutive relations. 
Table~\ref{table_vsi_param} shows that the decrements in the \tb{error} (F-values) are low for the filtered full-field data of prescribed displacement $\delta_1 =$ \SI{2.5}{\milli\meter}, suggesting that small displacement values do not generate enough non-linearity in the data sets to warrant multiple Ogden terms. 
Filtering the displacement also suppresses fluctuations in the predicted effective small strain parameters. 

\begin{table}[b]
\centering
\caption{\textbf{Effective small strain moduli estimated as averaged quantities from table~\ref{table_vsi_param}.} \tb{Additionally, a percentage difference of the small strain modulus with the best fit from the validation set is shown on the far right. All moduli are in \SI{}{\kilo\pascal}.}}
\label{table_small_strain}
\begin{tabular}{c cccc}\hline
\textbf{Specimen}  & $\kappa$ & $\mu$ & $\nu$ & $\%$\text{ diff. in }$\mu$     \\\hline \hline
\textit{Holes (Raw)}  & $6649.66\pm 1588.28$	& $17.06\pm 2.54$ &$0.4987\pm4.97\times10^{-4}$ & 7.67\\
\textit{Holes (Filtered)}  & $2413.34\pm 0.69$	& $16.47\pm0.29$	&$0.4966\pm2.32\times10^{-4}$ & 4.15\\\hline
\end{tabular}
\end{table}

\begin{figure}[t!]
    \includegraphics[width=\textwidth]{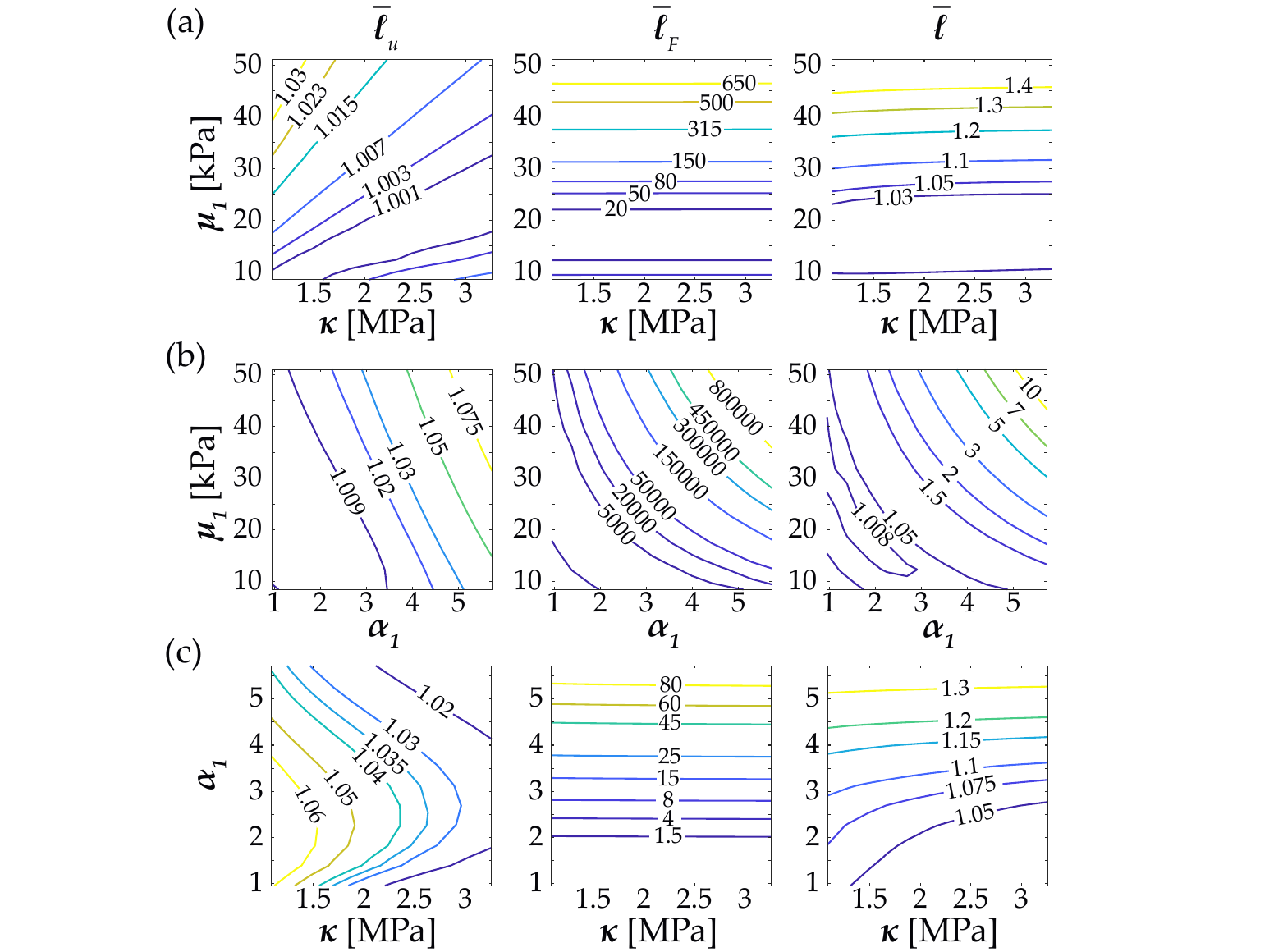}
     \caption{
    \textbf{Contours of the ratio of normalised \tb{error} components for displacement data \tb{($\overline{\ell}_u$)} and load data \tb{($\overline{\ell}_F$)}, and the normalised total \tb{error} ($\ell$) with sampling over $\kappa, \mu \text{ and } \alpha$.} The normalisation is done with respect to respective \tb{error} values evaluated at the parameters that minimise total \tb{error metric}. The \tb{errors} are computed for filtered data with $\delta_1 =$ \SI{2.5}{\milli\meter}.  
     }
    \label{fig:VSI_uncertainty}
\end{figure}

We investigate the local behaviour of the predicted model in terms of the change in \tb{error} due to a perturbation in the inferred parameters. 
The parameters are sampled around the trained 1-term model for the filtered data of the $\delta_1=$ \SI{5}{\milli\meter} case and the \tb{error} terms are evaluated for the filtered data of the $\delta_1=$ \SI{2.5}{\milli\meter} case. 
These normalised \tb{error} terms are defined by first estimating $\xi_0=\{\kappa_o, \mu_o, \alpha_o\} = \arg\min (\ell)$ from sampled points and then considering the \tb{displacement data error} $\overline{\ell}_u = \ell_u/ \ell_u|_{\kappa_o, \mu_o, \alpha_o}$, \tb{force error} $\overline{\ell}_F = \ell_F/ \ell_F|_{\kappa_o, \mu_o, \alpha_o}$, and \tb{normalised mean error} $\overline{\ell} = \ell/ \ell|_{\kappa_o, \mu_o, \alpha_o}$.
\tb{Error metric} contours for the normalised \tb{error} terms are presented in figure~\ref{fig:VSI_uncertainty}.
Since the material is shown to be near-incompressible, and thus volume preserving, the contour plots suggest that the bulk modulus determined by our optimisation scheme is insensitive to the model's behaviour. 
This is further confirmed by our observation that $\overline{\ell}$ does not change with $\kappa$, suggesting that the shear modulus determines the traction on the surface.

\tb{The uniaxial tension validation data, subject to the constraint of including a Neo-Hookean-like term to describe the small-strain behaviour, and the VSI method are similar and produce only marginal differences in material property estimates (table}~\ref{table_small_strain}).
These differences become more pronounced as constraints are relaxed. 
A driving factor behind these differences is the inherent non-uniqueness and covariance of the Ogden model with two and three terms~\cite{Jones2018}. 
All models are mathematically equivalent, stable, and produce a good fit to the experimental data, yet their parameters, in general, may differ considerably. 
This is predominantly a limitation of using a single uniaxial fit alone; improvements on reducing the likelihood of non-uniqueness and covariance in the fit model could be performed by incorporating more testing methods, such as biaxial tension, tension-torsion, etc. as in the classical work by Treloar~\cite{Treloar1943}.

\tb{It is worth noting that in system identification techniques, we typically start with a large basis, sequentially eliminate the operators with the least contributions, and retrain the model with the smaller basis.
This process is repeated until a substantial increase in error is seen.
The operators that are left at the end represent the different deformation mechanisms of the material. 
However, in the case of the Ogden model, all the operators have the same functional form. 
Therefore, we take the alternative approach of adding Ogden branches until the error plateaus. 
This works especially well with PDE-constrained minimisation which is very efficient for smaller set of parameters (typically 3 to 7 in our case studies).}

\section{Conclusion}

In this work, we present a methodology for leveraging fully three-dimensional deformations acquired by MR-\textit{\textbf{u}} and axial load data, establishing experimental sensitivity to a parameter, and using variational methods to not only identify Ogden material parameters from a single load step, but to do so with parsimony. 
Specific considerations arise from the choice of the Ogden model, which contribute challenges in interpretability and stability of fits. 
We draw several important conclusions from the work, and expect to pursue many of them in following studies. 
We note that filtering of experimental data continues to be important; while the small strain behaviour of our shear samples agrees for both unfiltered and filtered displacement data, challenges in convergence and expected values of the exponent parameters $\alpha_i$ illustrate this point. 
As perhaps was to be expected, increasing the scalar stretch amplitude $\lambda$ improves the observability of all material parameters by the integrated experimental sensitivity metric. 
Including deliberate heterogeneity in the displacement fields via inclusion of holes thus increased the number of voxels with larger stretch magnitudes (i.e. those near surfaces), and we observed improvements in both the sensitivity metric values and our numerical fits. 

As the decomposition and sensitivity assessment procedure is general, we anticipate its use for more complicated hyperelastic material behaviour \tb{for e.g. biomaterials. 
A potential path forward in the sensitivity analysis would be to use the physical, anisotropic invariants of Criscione et al.~\cite{Criscione2001}, but instead leveraging work functions of the deformation gradient tensor.  
In tandem with e.g. fiber direction fields derived from MRI as in \textit{in vivo}}~\cite{Wilson2021} \tb{or \textit{ex vivo}}~\cite{Luetkemeyer2021} \tb{techniques, we envision coupling of multimodal information and VSI for material versus structural~\cite{SplayPaper} characterisation.}
\tb{Another potential path forward} includes enhancing the observability of all material parameters in experiments based on procedurally tuning test geometry. 
New developments in machine learning, topology optimisation, or another form of constrained optimisation could be straightforwardly leveraged to this end, and (we hope) perhaps the next 50 years of celebrating this illustrious model.


\subsection*{Acknowledgements:} 
The authors would like to acknowledge Prof. Alan Wineman for impactful conversations on early versions of the theory.

\subsection*{Funding:}
Mechanical Engineering Department at the University of Michigan Startup Fund (JBE)\\
Rackham Graduate Fellowship (DPN)\\
National Science Foundation DMREF grant \#1729166 (SS, KG)

\subsection*{Author Contributions:}
Conceptualization: EMA, KG, JBE\\
Methodology: DPN, SS, BAA, UMS, JBE\\
Investigation: DPN, SS, UMS\\
Software: DPN, SS, BAA, JBE\\
Visualization: DPN, SS, JBE\\
Supervision: JBE, KG\\
Writing--original draft: DPN, SS, BAA, JBE\\
Writing--review \& editing: DPN, SS, BAA, UMS, EMA, KG, JBE

\subsection*{Competing Interests:}
The authors declare no competing financial interests or personal conflicts of interest that could have influenced the work described in this paper.

\end{document}